# Tools and Techniques for Managing Clusters for SciDAC Lattice QCD at Fermilab

A. Singh, D. Holmgren, R. Rechenmacher, S. Epsteyn
Fermi National Accelerator Laboratory,
Batavia, IL 60510, USA

## 1. Introduction

At Fermilab we have two SciDAC (Scientific Discovery through Advanced Computing) funded linux clusters, a 48-node and 128-node Xeon cluster[1] and an 80-node Pentium III cluster. We have minimal manpower available for administrating these clusters. We anticipate growth in our cluster facility to order 1000 nodes over the next few years. We have written tools and developed techniques that enable us to remotely administrate and automate repeated tasks to monitor these clusters and keep them in production. These are divided into 3 sections.

1. Hardware management tasks such as remotely powering on, powering off and power cycling a node or a group of nodes.
2. Operating system installation and upgrades, involving a large set of nodes, over Ethernet. This also includes reloading the BIOS or firmware on a set of nodes.
3. A set of tools that manipulate OpenPBS (Portable Batch System) [1] command outputs, allowing integration with our own parallel command processing and file copying tools.

This paper discusses how we integrate all these tools and techniques together to administrate and monitor the clusters to keep them in production. Henceforth in the paper we will refer to the server as the *head* node and the clients as the *worker* nodes.

## 2. Remote node management

Each of the worker nodes is installed with a BMC (Baseboard Management Controller). The BMC manages the interface between system management software and the platform management hardware, provides autonomous monitoring, event logging, and recovery control.

IPMI (Intelligent Platform Management Interface) [2] uses a message-based protocol for the different interfaces to the platform management subsystem such as serial/modem, LAN, PCI Management Bus, and the system software-side "System Interface" to the BMC.

The key characteristic of IPMI is that inventory, monitoring, logging and recovery is available independent of the main processors, BIOS and operating system. IPMI provides a standard protocol, which allows the user to format data packets with defined headers and payload to communicate with various data sensor repositories in the BMC on the motherboard via a network interface, serial link, or internal I/O port. Platform management functions are also available when the operating system is not running, or when the computer is in a powered-down state.

We have serial links connecting each worker node to a serial mutiplexer switch, which in turn connects via a SCSI type cable to a PCI card on the console server. This can be visualized as having the console server connected via a serial link to 48 or 128 nodes at one time. This allows us to redirect BIOS/console output from each worker node onto the console server and also allows us to power on, power off or power cycle a single node or a group of worker nodes. We support both IPMI versions 0.9 and 1.5. The only significant difference in these two versions with respect to our software is in the layout of the sensor data structures.

We also use the GUI-based cluster management application *IPMIView*, a SuperMicro product written in Java, which allows us to communicate with the BMC over Ethernet. It allows us to power on, power off and power cycle a single worker node and check on health (CPU temperature, fan speed, voltages etc.). The software has console redirection along with other remote node management options. We also have written a command-line utility with functions similar to *IPMIView,* which allows us to communicate with the BMC on a worker node via LAN. This is useful for scripting, for example, when the same command needs to be executed on more than one node, a feature lacking in the GUI-based software.

## 3. Network boot

PXE (Pre-eXecution boot Environment) code uses DHCP or BOOTP, in cooperation with a network server, to dynamically generate an IP address for the client node. This enables the PC to establish an IP connection with the server before the local operating system loads.

When PXE starts running, it looks for a BOOTP server. It is provided by the BOOTP server with various information, such as IP address, the DNS name server, the file server from which it should request a boot file, and the path to that file. Once the client receives the required information it loads the bootloader *pxelinux* for an OS install, or *pxegrub* for a BIOS or firmware install or upgrade. We have modified *pxegrub* so that it can load a DOS image. The bootloader loads either a linux kernel image and a linux ram disk, or a DOS image prepared with the mknbi utility [3].

---

[1] http://lqcd.fnal.gov

The bootloader is the program which first gets control of the machine from PXE. It first initializes and manages the raw hardware. Its primary job is to copy the operating system into host memory and pass control to the operating system. The bootloader performs any necessary platform specific hardware configuration for the operating system.

The initial ram disk provides the capability to load a Linux file system during the boot process. This ram disk can then be mounted as the root file system and programs can be run from it. Afterwards, a new root file system can be mounted from a different device. The previous root (from the initial ram disk) is then moved to a directory and can be subsequently un-mounted. The contents of the initial ram disk are designed to allow system startup to occur in two phases, where the kernel comes up with a minimum set of compiled-in drivers, and where additional modules are loaded from the new root file system.

For an operating system install (Linux) the following steps are executed once the kernel and ramdisk have been loaded:

- via init/rc.d, bring in an install script, which
    - partitions the disk.
    - makes a file system.
    - rcp's tar files for the file system.
    - unpacks the tar files.
    - does host IP configuration.
    - rsh's to the BOOTP server to comment itself out of the /etc/bootptab file.
    - reboots.

For a BIOS or firmware install or upgrade the following steps are executed once DOS and the ramdisk have been loaded:

- autoexec.bat executes, which then
    - does the firmware/BIOS upgrade/install.
    - using BOOTP discovers the host IP.
    - ftp's stdout/stderr to bootp server.
    - rsh's to bootp server to comment itself out of the /etc/bootptab file.
    - reboots.

The above procedure can be executed in parallel for more than one worker node. Since the procedure involves a single head node (bootp server) serving all the worker nodes, it involves a single point of failure. To avoid such a failure we execute the procedure on a fixed number of worker nodes, chosen depending on the network transfer rate and the maximum load bearing capacity of the bootp server. The above procedure does not scale well to hundreds of worker nodes so a multicast version is under development.

## 4. Fermi Tools

In this section we will talk about three primary tools which we use either standalone or in conjunction with each other. They are *rgang*, *fermistat* and *fermitrack*.

Nearly every system administrator tasked with operating a cluster of Unix machines will eventually find or write a tool, which will execute the same command on all of the nodes. At Fermilab we call this tool *rgang*. On each node rgang executes the given command via rsh or ssh, displaying the result delimited by a node-specific header. The original *rgang* at FermiLab was implemented in Bourne shell.

Because the original *rgang* executed the commands on the specified nodes serially, execution time was proportional to the number of nodes. We have implemented, in Python, a parallel version of *rgang*. This version forks separate rsh/ssh children, which execute in parallel. After successfully waiting on returns from each child or after timing out, this version of *rgang* displays the node responses in identical fashion to the original shell version of rgang. In addition, the latest *rgang* returns the OR of all of the exit status values of the commands executed on each of the nodes.

Simple commands execute via this *rgang* on all 80 nodes of one of our clusters in about 3 seconds. To allow scaling to kiloclusters, the new *rgang* can optionally utilize a tree-structure, via an "nway" switch. When so invoked, *rgang* uses rsh/ssh to spawn copies of itself on multiple nodes. These copies in turn spawn additional copies. A few examples below explain this in more detail.

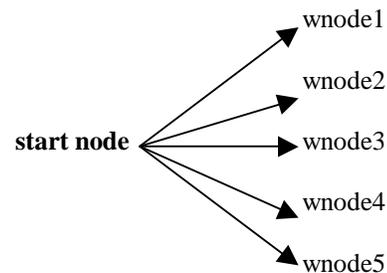

Example 1. nway = 0 (default)

In example 1 with an nway option of 0 or default, the *rgang* command forks off rsh or ssh commands in parallel from the start node onto all the worker nodes.

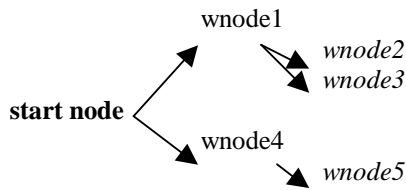

Example 2. nway = 2

In example 2 with an nway option of 2, the *rgang* command uses rsh/ssh to spawn copies of itself onto multiple nodes. These copies in turn spawn additional copies.

What motivated us to write *fermistat* was the desire for an interface that would integrate tools from OpenPBS with *rgang*. In the PBS execution environment a worker node is in one of the following states at any time: free, down, offline, reserved, job-exclusive or job-sharing. PBS provides commands that allow a system administrator to set a worker node free or offline depending on the state of the node. *Fermistat* works in conjunction with the PBS tools, taking a list of worker nodes as generated by *rgang* and executing the PBS command in series to set the appropriate state of the worker node.

For example, should the worker nodes wnode21, wnode22, wnode23 and wnode24 have a hardware problem and need to be placed offline for further investigation, we could execute the following commands in the PBS environment:

    [me]$> pbsnodes –o wnode21
    [me]$> pbsnodes –o wnode22
    [me]$> pbsnodes –o wnode23
    [me]$> pbsnodes –o wnode24

With *fermistat* we can affect the same result in a single command as follows.

    [me]$> fermistat –o wnode2{1-4}

The pattern wnode2{1-4} is passed to *rgang*, which expands the pattern to a list of worker nodes as wnode21, wnode22, wnode23 and wnode24. *fermistat* executes the *pbsnodes* command in series for each worker node in the expanded list. Scaling to hundreds of nodes is not very efficient as the above commands are executed in series. A parallel version of *fermistat* is under development. *rgang* can be used to generate node lists from more complex patterns than shown in the example above.

Consider a 4-node parallel job 12xy.myjob executing on the nodes wnode33, wnode53, wnode61 and wnode84. The *fermistat* "-l" option allows the user to list the nodes belonging to a PBS job as follows:

    [me]$> fermistat –l 12xy.myjob
    wnode33
    wnode53
    wnode61
    wnode84

This output can be piped to *rgang* to execute the desired command as follows:

[me]$> fermistat –l 12xy.myjob | rgang - <command>

The list of nodes generated by *fermistat* can be piped to another invocation of *fermistat* to put the nodes offline as follows:

    [me]$> fermistat –l 12xy.myjob | fermistat -o -
    pbsnodes –o wnode33
    pbsnodes –o wnode53
    pbsnodes –o wnode61
    pbsnodes –o wnode84

*Fermitrack* is a primitive accounting system, used in conjunction with the OpenPBS accounting system. OpenPBS appends accounting information such as resource usage (wall time, cputime, node usage, and so forth) into an accounting file. The command to submit a pbs job (*qsub*), patched with code from Argonne National Lab [4], checks a flat project file for either the default project associated with the current user, or for a valid project specified by the user. If neither succeeds the job is rejected. *fermitrack* reads accounting information each night from the OpenPBS accounting file and charge projects for cluster usage. If a particular project exceeds its limit, *fermitrack* removes it from the project file, thus automatically ensuring that a future *qsub* will reject that project when a job is submitted under that project name. *fermitrack* also keeps an account of cluster usage per project for later investigation and charge-back.

## 5. Integration of tools for health and status monitoring

Monitoring, logging and health data are transferred at regular intervals to the head nodes of our clusters. The worker nodes collect health data (CPU temperature, fan speed, voltages etc) from the sensor data repository maintained by the BMC using our IPMI software and send this information to the head node via the *syslog* udp socket.

At regular intervals, the head node issues a series of commands via *rgang* to check on resource usage on each worker node, such as disk space and client services (such as the OpenPBS client process). The head node checks whether the resource usage and health data for each worker are within safe limits. If not, alarms are generated in the form of an email and a blinking node name on the web monitoring pages.

A perl script that produces a graphical representation displaying the monitoring data in an easy to comprehend format for both the system administrator and user generates the web monitoring pages. The status of each node is displayed in a color-coded format. Nodes that have health or resource problems are highlighted in red to attract attention.

The head node also executes *fermitrack*, our software accounting extension to OpenPBS, at regular intervals to check on project resource usage. In case of project violations, when all assigned time is used up by a particular project, an email is sent to notify the system administrators to take appropriate action.

## 6. Conclusion

The tools and techniques that we have coded and developed at Fermilab for our SciDAC Lattice QCD clusters provide us flexibility and hardware independence for monitoring and debugging. They allow us to add or remove extensions to current available open source monitoring software, to integrate our own tools with current open source software, and to develop more scalable monitoring software to meet our growing demands.

The primary limitations to our current monitoring tools are single points of failure and scalability. As we add faster nodes and computer networks to our current setup, we expect execution times to reduce, thus allowing some scaling to larger numbers of workers. Another solution is to develop multicast versions of our tools. Still another is to have more than one monitoring node collecting health and resource usage data for groups of nodes, thus eliminating the single point of failure.

As we scale to many hundreds of nodes in our clusters, monitoring and administration require more manpower, attention and capacity. This growth will continue to drive development of new tools and techniques.

## Acknowledgments

Work supported by Universities Research Association Inc. under Contract No. DE-AC02-76CH03000 with the United States Department of Energy.